\documentstyle[12pt]{article}
\begin{document}
\begin{center}
\bf{DIQUARK STRUCTURE IN HEAVY QUARK BARYONS IN A GEOMETRIC MODEL}
\end{center}
\vskip 1 cm
\begin{center}
{\bf Lina Paria and Afsar Abbas} \\
Institute Of Physics \\
Bhubaneswar-751005 \\
India \\
e-mail : lina@iopb.ernet.in \\
e-mail:  afsar@iopb.ernet.in  
\vskip 1 cm
\bf{Abstract}
\end{center}
Using a geometric model for the study of the structure of
hadrons, we
study baryons having one, two and  three heavy quarks. The study 
reveals diquark structure in baryons with one and two heavy quarks
but not with three heavy  identical quarks.
\pagestyle{plain}
\newpage
The study of the heavy quark systems \cite{ar1} is an important issue
in particle physics. It has even become more so in the recent
years \cite{ar2}
 as several existing and planned machines are expected to produce a barrage
 of experimental information in the near future. The field has
received a boost from the so called Heavy Quark Effective Theory
\cite{isgur,ar3,neub}. An interesting property which arises in this theory
is the existence of the diquark structure for two heavy quarks (QQ) in
a baryon consisting of $ QQq $, (where $ q $ is for a light quark)
\cite{savag,falk}.

Now the study of diquark is almost as old as that of the quarks. Existance
or nonexistance of diquark structure in baryons with (i) $ qqq $, 
(ii)  $  qqQ $, (iii) $ qQQ $, (iv) $ QQQ $ has been a problem of much interest
\cite{rich,licht} and has recently been reviewed \cite{lich}. Most studies
have been done within the framework of potential models,
bag models and string models \cite{lich}. The cases (i), (ii) and (iii)
have been well studied but not the case (iv). In our study in addition
to the study of the cases (i), (ii), (iii) we shall do case (iv)  as
 well. However the information extracted as to the diquark structures 
obtained by different people are often in direct
conflict with each other. Some consensus has been achieved but there is
much confusion. The situation can be best summerised by quoting the 
statement of the "five" authors  of the review article  \cite{lich}
".....sometimes we do not agree  among ourselves about the nature 
of diquarks".

Our aim in this paper is to study this important dual problem of the heavy
quarks and of the diquark structure in baryon within a framework which is
complementary to the potential models and the bag models pictures. This
is a geometric model made use of recently \cite{iache,halse} in the
context of light baryons. In fact a diquark structure was obtained
 therein  \cite{halse}. This view complements and supports
the view that the nucleon is deformed in the ground state. This
was  obtained
within the configuration  mixed wave function  picture in a quark potential 
model \cite{aa}.

In the geometric model of hadrons quarks sit at  different positions
in a collective picture \cite{iache,halse}. All the excited states of
hadrons are obtained by rotations and vibrations of quarks in a collective
mode. For simplicity, we will be considering only the rotational
 excitations.
 The total wave function of baryons is given by 
 \begin{equation}
  \psi_{total} = \psi_{c} \otimes \psi_{sf} \otimes \psi_{r} 
 \end{equation}
 where $ \psi_{c} $, $ \psi_{sf} $ and $ \psi_{r} $ stands for the
wave function corresponding to color, spin-flavor and the geometric
degrees of freedom respectively. 
From the antisymetrisation of the color state, the product of the wave
function $ \psi_{sf} \otimes \psi_{r} $  has to be totally symmetric.

To understand the geometric structure of baryons, we are considering
the three possible geometrical arrangements of quarks as in 
Fig. 1. Here \\
general case  (a) : two quarks are seperated by a distance $ r_{2} $ ,
while the third quark is at a distance $ r_{1} $ from the center of mass
of the first two quarks. \\
case (b) : $ r_{2}=0 $ ;  i.e. two quarks are at the same place while the
third one is separated.
case (c) : $ \frac{r_{1}}{r_{2}} =\frac{\sqrt{3}}{2} $ ;  i.e. 
three quarks are sitting at the vertices
of an equilateral triangle.
case (d) : $ r_{1} =0 $ ;  i.e. all quarks are equidistant and lying
on a line.
The point group structure for inertia tensor ( $ G_{I} $ ), for 
equivalent particles ( $ G_{E} $ )  and  for distinct particles
( $ G_{D} $ ) are given below respectively.
\begin{eqnarray}
general ~~  case ~~ (a)  & : &  C_{2  v }  ,  C_{2  v } ,  S_{1}
\nonumber \\  
case  (b) & : &  C_{\infty v}  ,  C_{\infty  v} ,  C_{\infty  v}
\nonumber \\   
case (c) & : &  D_{3 h} ,  D_{3 h} ,  S_{1} \nonumber \\
case (d) & : & D_{\infty h} ,  D_{\infty  h } , C_{\infty v}  
\end{eqnarray}      
As we are considering baryons with a heavy flavour ( charm)  the
corresponding spin flavor group symmetry is $ SU_{sf}(8) $. 
The group  $ SU_{sf}(8) $ is broken in the following chain of 
subalgebras 
\begin{eqnarray}
SU_{sf}(8) & \supset & SU_{f}(4) \otimes SU_{s}(2) \nonumber \\
& \supset & SU_{f}(3) \otimes 
U_{c}(1) \otimes SU_{s}(2) \nonumber \\
 & \supset & SU_{I}(2) \otimes U_{Y}(1)\otimes
U_{c}(1)\otimes SU_{s}(2) \nonumber \\
& \supset & SO_{I}(2)\otimes U_{Y}(1) \otimes
U_{c}(1)\otimes SO_{s}(2)
\end{eqnarray}
The decomposition of the  relevant representations of 
$ SU_{sf}(8) $ into \\ $ SU_{f}(4) \otimes
SU_{s}(2) $ are:
\begin{eqnarray}
120 & \supset & {^4}{20} \oplus {^2}{20} \nonumber \\
168 & \supset & {^2}{20} \oplus {^4}{20} \oplus {^2}{20} \oplus{^2}{\bar{4}}
\nonumber \\
56 & \supset & {^2}{20} \oplus {^4}{\bar{4}} 
\end{eqnarray}
The  quantum numbers of the baryons $ N $, $\Delta $,
${\Lambda}_{c} ,  {\Sigma}_{c} ,   {\Xi}_{cc} ,  {\Omega}_{ccc} $
under $ SU_{sf}(8) $ group symmetry are given in Table I.
Fig. 2 is the schematic diagram  drawn on the basis of  table 2 in ref
\cite{halse}. This shows the variation of levels with a change
in the shape of the baryons (as stated in  the cases ((b),(c),(d)))
for $ L=0 $   \&  $ L=1 $.  Here $ L $ is the orbital angular momentum
and $ K $ is the projection of the orbital angular momentum on the
body fixed axis with $ L = K,  K+1,  K+2, ..... $. The notation used 
here is :  $ K = 0+ $ indicate $ L = 0, 2, 4, ..... $ and
$ K = 0- $ indicate $ L = 1, 3, 5, ..... $. 
This diagram is a slightly modified
version of the one by Halse \cite{halse}.

Note that the spectrum generating algebra chain in equation (3) implies
that the symmetry is broken diagonally. ( It is because of this, that
the fact that the c- quark is much heavier than the u-, d-, and s- quarks 
is not expected to effect our analysis ). This means that the energy 
levels  may depend on the eigen value of the Casimir operators of
the group chain. Hence the $ SU(8) $ mass formula is
 \begin{eqnarray}
 M^{2} & = & M_{0}^{2} + a[\hat{C_{2}}(SU(8))-\frac{231}{8}]
\nonumber \\ & + & b[\hat{C_{2}}(SU(4))-  \frac{39}{4}]  
  +   c[\hat{C_{2}}(SU(3))-\frac{8}{3}]
 +d[s(s+1)-\frac{3}{4}] \nonumber \\ 
 & + &   e[C-1] + f[I(I+1)] + \alpha  L  + \beta  K 
 \end{eqnarray}
 \noindent
 where  $ \hat{C_{2}} $  is the eigenvalue of the Casimir operator
 for different representation, $ C $ is charm quantum number, $ s $ is
spin value, $ I $ is isospin, and $ a $, $ b $, $ c $, $ d $, 
$ e $, $ f $,
$ \alpha $ , $ \beta $ are parameters. The term $ \alpha L $ and
$ \beta K $  in the mass formula comes from the rotation of the system.
The operators are defined in such
a way that for the ground state of $ \Lambda_{c} $  all the
terms  except $ M_{0}^{2} $  vanish. The justification for using
the linear term in $L$ in the above formula arises from the fact
that our model is intrinsically related to the stringlike bag model
(as evident from  Fig. 1 ). This has also been pointed out
in ref.[11].

We treat Kalman and Tran's result \cite{kal} on  the heavy quark baryons
as  a good representative sample of the theoretical studies in this area.
In addition these results are published in a great detail making them
suitable as a point of reference. We will treat these numbers as "experimental
numbers".  Note that we do not consider $ \Xi_{c} $ , $ \Omega_{c} $,
$ \Omega_{cc} $ etc. as the corresponding data does not exist. 
We do not expect much changes in our basic conclusions when
the real experimental numbers become available. The  available 
experimental data is from  \cite{pdg}.

For each baryons, we arrange the states in such a way that the energy
will increase from lower excited states to higher excited states. Knowing
the internal quantum numbers, spin-flavour symmetry of each baryon 
considered  is given in Table I . Knowing the Casimir operator 
value of the relevant representations
and using the mass formula we can specify each states of these baryons
in a particular $ [ g ] L ^{\pi} $ representation. Here ${\pi}$ is the
parity of the state and $ [ g ] $ is the $ SU(8) $ representation in 
the standard Young Diagram  formalism giving the total 
number of boxes in each row, e.g.  $ [ 2 1 ] $ denote the two
boxes in first row and one box in second row.
The notation $ (p_{1}  p_{2}  p_{3} ) $ in Table I denote the difference
in the number of boxes between the rows. 
i.e   $ p_{i} = {\lambda}_{i} - {\lambda}_{i+1} $,   where ${\lambda}_{i}$ is
the number of boxes in the $ i ^{th} $ row of  the Y.D.  
 Our approach for assignment of the states of baryons is that 
of Halse \cite{halse} where the mass formula is used as a guide 
for these assignments.

As an example, for $ \Lambda_{c} $ :  \\
(i) The lowest energy state $ J^{\pi} (M) $ is $ \frac{1}{2}^{+}(2282) $,
with $ L = 0 $, $ s = \frac{1}{2} $, $ K = 0 $, \\
( 1 1 0 ) representation
of $ SU(4) $ , ( 0 1 ) representation of $ SU(3) $. Now using the Casimir
operator value of each representation in  the mass formula we get 
$ M^{2} = M_{0}^{2} $ and the state $ \frac{1}{2}^{+}(2282) $ goes
to the $ [g] L^{\pi} $ representation as $ [3] 0^{+} $. \\
(ii) The next excited state $ J^{\pi} (M) $ is $ \frac{1}{2}^{-}(2653) $,
with  $ L = 1 $, $ s = \frac{1}{2} $, $ K = 0 $, \\
( 1 1 0 ) representation
of $ SU(4) $ , ( 0 1 ) representation of $ SU(3) $. This assigns this state
in the  $ [ g ] L^{\pi} $ representation as
$ [3] 1^{-} $ ( which gives a reasonable  fit ) rather than the assignment
 $ [ 2 1 ] 1^{-} $ and so on. In this way the states of each baryons
considered in Table I are assigned  and a global fit of the states are
able to give the parameter values as :  $ M_{0}^{2} = 6.303 $,
$ a = -0.317 $, $ b = 0.163 $, $ d = -0.152 $, $ \alpha = 1.297 $,
$ \beta = 1.022 $, $ c = 9.5025 $, $ e = 18.7795 $, $ f = -18.994 $ .
 All parameters are in unit  $GeV^{2}$.

The low excited states of  $ N $, $ \Delta $, $ {\Lambda}_{c} $ , 
$ {\Sigma}_{c} $, $ {\Xi}_{c  c} $ , $ {\Omega}_{c  c  c} $  
have been given the assignment shown in Table II. The location of
these particles as per our assignment is also indicated in Fig. 2.

We find that the nucleon has the structure falling
in between the case(b) and (c) ( Fig. 2.).
The position of $ \Delta $ is  found to be slightly  to the right of
$ N $ (see Fig. 2.). This is in contrast to  what Halse \cite{halse}
 had obtained for $ N $ and $ \Delta $. This is because 
in his fit he was trying to include a single star state 
$\frac{1}{2}^{+}(1550) $ which existed then (i.e. in the 1986 data set),
but does not exist anymore ( i.e. in the 1994 data set \cite{pdg} ).
Instead a new state $ \frac{1}{2}^{+}(1750) $  has arisen whose presence
makes the above difference. However still in aggrement with Halse \cite{halse},
 we  obtain a diquark structure in  $ \Delta $.

For $ {\Lambda}_{c} $  and  $ {\Sigma}_{c} $ we see that the order of the
representations is such that it's geometric structure tends to
move  towards the case (b) ( Fig. 2.).
This indicates that for the one heavy quark baryon,the diquark structure
exists ; i.e. two light quarks (qq) can form  a diquark in
(qq-Q) while  restoring the
$ C_{2 v} $ summetry. This is in agreement with the result of
Lichtenberg  \cite{licht}.
In the case of $ {\Xi}_{cc} $ baryons, there too is a diquark
structure.   So  one  can
say that in  (QQ-q), the two heavy quarks come together
to form a diquark. This view is in agreement with others
\cite{savag,falk,rich,lich}. Here we see that though the diquark
structure exists in both the baryons containing one and two heavy quarks,
but the nature of the diquark in these two cases is different due to the
$ C_{2 v} $ symmetry considerations. We are considering only the low
angular momentum states so the effect on the diquark structure due
to higher angular momentum is not being looked into here.

The three heavy quark baryon $ {\Omega}_{ccc} $ shows the structure of
that of the case(c) ( Fig. 2.). This is quite reasonable as 
all the three quarks are equivalent. This is a new interesting result 
as it appears that not much  work has been done by others in the 
three heavy quark case. So the conclusion of our model is that 
the diquark structure exists in one  and two heavy quark baryons 
but not in the three  identical heavy  quark baryons.

\vskip 2 true cm
The authors would like to thank the referee for usefull comments.

\vfill
\newpage
\centerline {\bf TABLE CAPTIONS }
\noindent {\bf Table I:}
$SU_f(3)$, $SU_f(4)$, $SU_{fs}(8)$ and $SU_s(2)$ levels for the
charmed and uncharmed baryons, classified according to isospin (I),
strangeness (S) and charm (C).
\vskip 1.0 true cm
\noindent {\bf Table II:}
Ordering of $[ g ]\ L^{\pi} $  ( the notation explained in 
the text ) representations in the assignment of
states for $N$, $\Delta$, $\Lambda_c$, $\Sigma_c$, $\Xi_{cc}$ and
$\Omega_{ccc}$. (Ground to excited states are ordered from top to
bottom.) 
\vfill
\newpage
\begin{figure}[p]

\centerline {\bf FIGURE CAPTIONS} 
\vskip 1.0 true cm

\caption{Geometrical arrangement of quarks separated by 
$  r_{1} $ , $ r_{2} $  in the general case  and three other
cases as discussed in the text.}

\caption{ Schematic variation of the levels $ [g]K $ 
with the change in the shape of the baryons. Note that $K = 0+ $ or $ 0-$
is explained in the text.( Warning: the symbols $ +/- $ are not
superscript and are not symbols for parity: see text).
The vertical axis is labelled by
the energy ( schematically) and the horizontal axis is labelled
by the values of $ r_{1} $ and $ r_{2} $ as  explained in  the text. 
The dashed lines labeled by
$ L^{\pi} $ for the negative parity levels and the solid lines
for the positive parity levels. The dotted line for the state 
 $ [3]0- $  is for the baryon containing at least one heavy quark (i.e. not
for $ N $ and $ {\Delta} $ ) . 
The arrows indicate the
location of different baryons as per our assignment.}
\end{figure}
\vfill
\newpage

\begin{table}
\centerline {\bf Table I} 
\vskip 0.2 in
\begin{center}
\begin{tabular}{|c|c|c|c|c|c|c|c|}
\hline
\multicolumn{1}{|c|}{Particle} &
\multicolumn{1}{|c|}{I} &
\multicolumn{1}{|c|}{S} &
\multicolumn{1}{|c|}{C} &
\multicolumn{1}{|c|}{$SU_f(3)$} &
\multicolumn{1}{|c|}{$SU_f(4)$} &
\multicolumn{1}{|c|}{$SU_{fs}(8)$} &
\multicolumn{1}{|c|}{$SU_s(2)$} \\
\multicolumn{1}{|c|}{} &
\multicolumn{1}{|c|}{} &
\multicolumn{1}{|c|}{} &
\multicolumn{1}{|c|}{} &
\multicolumn{1}{|c|}{$(p_1\ p_2)$} &
\multicolumn{1}{|c|}{$(p_1\ p_2\ p_3 )$} &
\multicolumn{1}{|c|}{$[g]$} &
\multicolumn{1}{|c|}{$Spin$} \\
\hline
$N$ & ${1\over 2}$ & 0 & 0 & (1\ 1) & (1\ 1\ 0) & [3] & ${1\over 2}$ \\
 &  &  &  &  &  & [2\ 1] & ${1\over 2}$, ${3\over 2}$ \\
  &  &  &  &  &  & [1\ 1\ 1] & ${1\over 2}$ \\
\hline
$\Delta$ & ${3\over 2}$ & 0 & 0 & (3\ 0) & (3\ 0\ 0) & [3] & ${3\over 2}$ \\
 &  &  &  &  &  & [2\ 1] & ${1\over 2}$ \\
\hline
$\Lambda_c$ & {0} & 0 & 1 & (0\ 1) & (1\ 1\ 0) & [3] & ${1\over 2}$ \\
 &  &  &  &  &  & [2\ 1] & ${1\over 2}$, ${3\over 2}$ \\
  &  &  &  &  &  & [1\ 1\ 1] & ${1\over 2}$ \\
 &  &  &  & (0\ 1) & (0\ 0\ 1) & [2\ 1] & ${1\over 2}$ \\
  &  &  &  &  &  & [1\ 1\ 1] & ${3\over 2}$ \\
\hline
$\Sigma_c$ & {1} & 0 & 1 & (2\ 0) & (3\ 0\ 0) & [3] & ${3\over 2}$ \\
 &  &  &  &  &  & [2\ 1] & ${1\over 2}$ \\
  &  &  &  & (2\ 0) &(1\ 1\ 0)  & [3] & ${1\over 2}$ \\
 &  &  &  &  &  & [2\ 1] & ${1\over 2}$,${3\over 2}$ \\
  &  &  &  &  &  & [1\ 1\ 1] & ${1\over 2}$ \\
\hline
$\Xi_{cc}$ & ${1\over 2}$ & 0 & 2 & (1\ 0) & (3\ 0\ 0) & [3] & ${3\over 2}$ \\
 &  &  &  &  &  & [2\ 1] & ${1\over 2}$ \\
  &  &  &  & (1\ 0) &(1\ 1\ 0)  & [3] & ${1\over 2}$ \\
 &  &  &  &  &  & [2\ 1] & ${1\over 2}$,${3\over 2}$ \\
  &  &  &  &  &  & [1\ 1\ 1] & ${1\over 2}$ \\
\hline
$\Omega_{ccc}$ & ${0}$ & 0 & 3 & (0\ 0) & (3\ 0\ 0) & [3] & ${3\over 2}$ \\
 &  &  &  &  &  & [2\ 1] & ${1\over 2}$ \\
\hline
\end{tabular}
\end{center}
\end{table}

\newpage
\begin{table}
\centerline {\bf Table II} 
\vskip 0.2 in
\begin{center}
\begin{tabular}{|c|c|c|c|c|c|}
\hline
\multicolumn{1}{|c|}{$N$} &
\multicolumn{1}{|c|}{$\Delta$} &
\multicolumn{1}{|c|}{$\Lambda_c$} &
\multicolumn{1}{|c|}{$\Sigma_c$} &
\multicolumn{1}{|c|}{$\Xi_{cc}$} &
\multicolumn{1}{|c|}{$\Omega_{ccc}$}\\
\hline
[3] $0^+$ & [3] $0^+$ & [3] $0^+$ & [3] $0^+$ & [3] $0^+$ & [3]
$0^+$ \\ 
$[2 1]$ $0^+$ & [3] $2^+$ & [3] $1^-$ & [3] $1^-$ & [3] $1^-$ &
[3] $2^+$ \\  
$[2 1]$ $1^-$ ($ K=0$) & $[2 1]$ $1^-$ ($K=0$) & $[2 1]$ $0^+$ & 
$[2 1]$ $0^+$ &
$[2 1]$ $0^+$ & $[2 1]$ $1^-$ ($K=0$) \\
$[2 1]$ $2^+$ & $[2 1]$ $0^+$  & $[2 1]$ $1^-$ ($K=0$) & $[2 1]$ 
$1^-$ ($K=0$) &
$[2 1]$ $2^+$ & $[2 1]$ $1^-$ ($K=1$) \\
$[2 1]$ $1^-$ ($K=1$) & $[2 1]$ $1^-$ ($K=1$) & [3] $2^+$ & [3] $2^+$ & 
$[2 1]$
$1^-$ ($K=0$) & \\
& & $[2 1]$ $1^-$ ($K=1$) & $[2 1]$ $2^+$ & & \\
& & & $[2 1]$ $1^-$ ($K=1$) & & \\
\hline
\end{tabular}
\end{center}
\end{table}


\newpage
\begin{thebibliography}{99}
\bibitem{ar1}  W.  Kwong, J.L. Rosner and C. Quigg,
Ann.Rev.Nucl.Part.Sci. {\bf37}, 325(1987)
\bibitem{ar2} J.G. Koerner and H.W. Siebert,
Ann.Rev.Nucl.Part.Sci. {\bf41}, 511(1991)
\bibitem{isgur} N. Isgur and M.B. Wise,
Phys.Rev.Lett. {\bf66} (1991) 1130
\bibitem{ar3} B. Grinstein, Ann.Rev.Nucl.Part.Sci.
{\bf42}, 101(1992)
\bibitem{neub} M. Neubert, Phys.Rep  C  {\bf245}, 259(1994)
\bibitem{savag} M.J. Savage and M.B. Wise, Phys.Lett. B
{\bf248}, 177(1990)
\bibitem{falk} A.F. Falk, M. Luke, M.J. Savage and M.B. Wise,
Phys.Rev.  D {\bf49}, 555(1994).
\bibitem{rich} S. Fleck, B. Silvestre -Brac and J.M. Richard,
Phys.Rev.  D {\bf38}, 1519(1988).
\bibitem{licht} D.B.Lichtenberg, J.Phys.  G {\bf16} 1599(1990).
\bibitem{lich} M. Anselmino, E. Predazzi, S. Ekelin, S. Fredriksson
and D.B. Lichtenberg, Rev.Mod.Phys. {\bf65}, 1199(1993).
\bibitem{iache} F. Iachello, Phys.Rev.Lett. {\bf62}, 2440(1989).
\bibitem{halse} P. Halse, Phys.Lett. B  {\bf253} 9(1991).
\bibitem{aa} A. Abbas, J.Phys G {\bf18}, 89(1992)
\bibitem{kal}  C.S. Kalman and B. Tran, Il.Nuovo.Cim. A
{\bf102}, 835(1989).
\bibitem{pdg}  Review Of Particle Properties, Phys.Rev.D
{\bf50} (1994)1173.
\end{thebibliography}
\end{document}